\documentclass[aps,amsmath,amssymb,reprint,superscriptaddress,10pt,nofootinbib]{revtex4-2}

\usepackage[dvipsnames]{xcolor}
\usepackage{graphicx}
\usepackage{changepage} 
\usepackage{multirow} 
\usepackage{hyperref}
\usepackage{lipsum}
\usepackage{cleveref}
\usepackage{soul}
\usepackage{comment}
\usepackage{booktabs} 
\usepackage{siunitx}  
\definecolor{green}{rgb}{0.01, 0.75, 0.24}

\begin{document}

\title{Chemical potential differentials in the QCD phase diagram from heavy-ion isobar collisions
}

\author{Joaquin Grefa}
\affiliation{Center for Nuclear Research, Department of Physics, Kent State University, Kent, OH 44242, USA}
\affiliation{Department of Physics, University of Houston, Houston, TX 77204, USA}

\author{Chun Yue Tsang}
\affiliation{Center for Nuclear Research, Department of Physics, Kent State University, Kent, OH  44242, USA}

\author{Rajesh Kumar}
\affiliation{Center for Nuclear Research, Department of Physics, Kent State University, Kent, OH  44242, USA}
\affiliation{Department of Physics, MRPD Government College, Talwara, Punjab 144216, India}
 
\author{Veronica Dexheimer}
\affiliation{Center for Nuclear Research, Department of Physics, Kent State University, Kent, OH  44242, USA}

\author{Claudia Ratti}
 \affiliation{Department of Physics, University of Houston, Houston, TX 77204, USA}

\author{Zhangbu Xu}
\affiliation{Center for Nuclear Research, Department of Physics, Kent State University, Kent, OH  44242, USA}
\affiliation{Physics Department, Brookhaven National Laboratory, Upton, NY, 11973, USA}

\date{\today}

\begin{abstract}
Temperature and baryon, charge, and strangeness chemical potentials characterize QCD matter under extreme conditions. Differences between these chemical potentials and their ratios probe conserved-charge correlations and the system’s response in the multidimensional QCD phase diagram. We extract these quantities from STAR Ru+Ru and Zr+Zr isobar collisions using a Bayesian thermal analysis of hadron yields, which substantially reduces systematic uncertainties, and compare them with Taylor-expanded lattice-QCD and Chiral Mean Field model predictions. Isobar collisions thus emerge as a precision probe of four-dimensional QCD thermodynamics.
\end{abstract}

\maketitle

\section{Introduction}
\label{Section I}

Understanding strongly interacting matter under extreme conditions is a central goal of nuclear physics~\cite{achenbach2024present}. At high energies, Quantum Chromodynamics (QCD) predicts a transition from hadronic matter to a deconfined quark–gluon plasma. This transition has been extensively studied through ultra-relativistic heavy-ion collisions at the Relativistic Heavy Ion Collider (RHIC) and the Large Hadron Collider (LHC). 
Measurements of identified hadron particle spectra and yields at mid-rapidity provide access to chemical freeze-out conditions, which can be extracted via thermal model analyses \cite{STAR:2005gfr,Andronic:2017pug}.
In this work, we perform a Bayesian analysis of hadron yields using the statistical model THERMUS~\cite{Wheaton:2004qb}.
The resulting posterior distributions provide both the marginal probability densities for and the correlations among the chemical‑freeze‑out temperature $T_{\text{Chem}}$, the fireball radius $R$, the set of chemical potentials $\mu_i$, and the strangeness‑suppression factor $\gamma_s$. These joint posteriors map out experimentally constrained trajectories in the QCD phase diagram.

The relevant $\mu$'s correspond to the conserved quantities of the system: baryon chemical potential ($\mu_B$), related to net baryon number ($B$) conservation; electric charge chemical potential ($\mu_Q$), related to net electric charge ($Q$) conservation (hereafter ``charge"); and strangeness chemical potential ($\mu_S$), related to net strangeness ($S$) conservation. While these quantities can be studied independently \cite{STAR:2005gfr,Andronic:2017pug}, most model fits to experimental data impose simplifying assumptions, such as $\mu_Q=0$ or a fixed charge fraction $Y_Q=Q/B$ at midrapidity equal to the proton to baryon ratio $Z/A$ of the colliding nuclei ~\cite{STAR:2005gfr,STAR:2017sal,Andronic:2017pug}. As a result, the interdependence among the $\mu_i$ and their mutual response (expressed as constituent correlations of susceptibilities~\cite{Fu:2018swz,Pratt:2016lol,Pratt:2017oyf,Ding:2025jfz,Karsch:2022jwp}) and through differentials (such as $d\mu_B/d\mu_Q$ and $d\mu_S/d\mu_Q$) remain largely unexplored in dynamically evolving heavy-ion experiments.

Isobar collisions~\cite{STAR:2021mii} provide particularly interesting systems to access this physics. Ruthenium ($^{96}_{44}$Ru) and Zirconium ($^{96}_{40}$Zr) have the same mass number $A = 96$, but differ in their proton and neutron content, corresponding to $Y_Q=44/96 \approx0.46$ and $Y_Q= 40/96 \approx 0.42$, respectively, a difference of $\sim9\%$. Studying how nuclear systems respond to such deviations from isospin symmetry ($Y_Q = 0.5$) is not only essential for heavy-ion phenomenology, but also for understanding neutron stars, where $\beta$-equilibrium drives the charge fraction to values as low as $Y_Q < 0.1$ \cite{Pons:1998mm}. 

\begin{figure}[t!]
    \centering
    \includegraphics[width=0.485\textwidth]{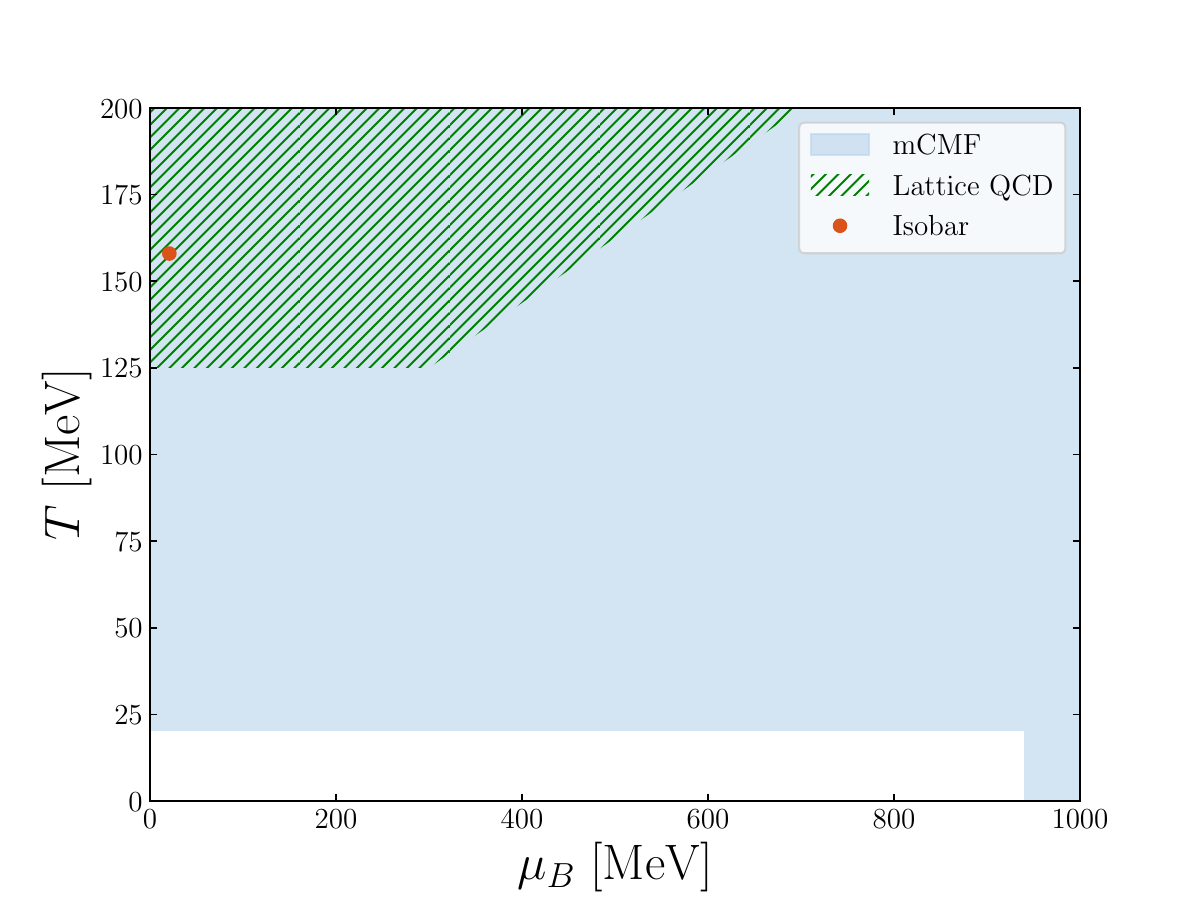}
    \caption{QCD phase diagram regions reproduced by experiment and covered by the theoretical descriptions discussed in this work.}
    \label{fig:QCDPD}
\end{figure}

From the theoretical side, first-principle calculations are limited to specific regions of the QCD phase diagram~\cite{MUSES:2023hyz}. Lattice-QCD calculations are currently limited to vanishing $\mu_B$ (due to the sign problem~\cite{Nagata:2021ugx}) and temperatures $T \gtrsim 110$ MeV. Lattice results can be extended to finite $\mu_B$ via Taylor expansions of thermodynamic quantities around $\mu_B=\mu_Q=\mu_S=0$, using generalized susceptibilities computed on the lattice~\cite{Noronha-Hostler:2019ayj,Monnai:2019hkn,Monnai:2024pvy}. The resulting equation of state, denoted here as $BQS$ \cite{Noronha-Hostler:2019ayj,jahan_2025_14639785}, provides a controlled way to include the effects of finite $\mu_{B}$, while incorporating finite $\mu_Q$ and $\mu_S$, and being reliable up to $\mu_B/T \approx 2.5$. This allows experimental constraints of strangeness neutrality (enforced via $Y_S=S/B=0$) and fixed $Y_Q$ to be imposed, covering the region relevant for the isobar collisions studied in this work (see \Cref{fig:QCDPD}). Alternative expansion schemes can extrapolate the equation of state up to $\mu_{B}/T=3.5$ \cite{Borsanyi:2021sxv,Abuali:2025tbd}. Beyond these regions, uncertainties grow rapidly, and one must turn to a different approach.

Provided that the appropriate degrees of freedom, interactions, and symmetries are included, effective models can be applied to describe the entire multidimensional QCD phase diagram. We make use of the effective Chiral Mean Field Model with an improved meson description (mCMF), including the baryon octet and decuplet. CMF is a relativistic mean-field effective model that describes baryons as interacting via meson mean fields. It incorporates chiral symmetry restoration and deconfinement to quark matter at high energies, both becoming crossovers for high $T$/low $\mu_B$. mCMF goes to some extent beyond the mean-field approximation by including interactions for the thermal mesons \cite{Kumar:2025rxj}. These interactions are calculated through in-medium masses of pseudoscalar and vector mesons, obtained from the explicit chiral symmetry-breaking and vector interaction terms in the Lagrangian, respectively, prior to applying the mean-field approximation. This extension improves the description of thermodynamics at high $T$/low $\mu_B$ \cite{Kumar:2025rxj}, corresponding to the region of the isobar collisions studied in this work (\Cref{fig:QCDPD}), while retaining applicability to the high-$\mu_B$/low-$T$ region relevant for neutron-star matter~\cite{Kumar:2024owe}.

\section{Methodology and results}
\label{Section II}

\subsection{From heavy-ion collisions}
\label{Section II_HI}

Previous results~\cite{STAR:2005gfr,STAR:2017sal,Andronic:2017pug,Wheaton:2004qb} suggest that particle yields in ultra-relativistic heavy-ion collisions can be fitted under the assumption of thermal equilibrium (see Supplemental Material, \Cref{A}), with relative yields determined by $\mu$'s at $T_{\text{Chem}}$. However, when Ru+Ru and Zr+Zr systems are analyzed independently, the resulting differences in $\mu$'s are consistent with zero due to large uncertainties~\cite{Stankus:2006zn,STAR:2017sal,ALICE:2013mez}. To address this issue, Ref.~\cite{Lewis:2022arg} proposed a method for measuring net-charge from the particle yield differences ($\Delta Q$) between isobar systems using double ratios.
\begin{eqnarray}
&\Delta& Q = [(N_{\pi^+} + N_{K^+} + N_p) - (N_{\pi^-} + N_{K^-} + N_{\bar{p}})]_\text{Ru} \nonumber \\&-& [(N_{\pi^+} + N_{K^+} + N_p) - (N_{\pi^-} + N_{K^-} + N_{\bar{p}})]_\text{Zr}\,,
\end{eqnarray}
where $N_i$ represents mid-rapidity yields of particle species $i$. Under the assumption that $\Delta Q \ll N_i\ \forall i \in (\pi, K, p)$, a change of variable simplifies $\Delta Q$ to:
\begin{equation}
\Delta Q \approx N_{\pi}(R2_{\pi}-1) + N_{K}(R2_{K}-1) + N_{p}(R2_{p}-1)\,,
\label{2}
\end{equation}
where $R2_i = [(N_i/N_{\bar{i}})_\text{Ru}/(N_i/N_{\bar{i}})_\text{Zr}]$ are double ratios, and $\bar{i}$ denotes the corresponding antiparticle. 
This change of variables significantly reduces systematic uncertainties, as most of them cancel out in the double ratios. 
Using this approach, the STAR experiment has compared the ratio $\left<B\right>/\Delta Q$ against $A/\Delta Z$ in Ru+Ru and Zr+Zr collisions, where $\left<B\right>$ is the average net-baryon number. In 0-10\% most central collisions, $\left<B\right>/\Delta Q\times \Delta Z/A=1.84\pm 0.02 \text{(stat)}\pm0.09\text{(syst.)}\pm0.16$(feed-down), favoring the baryon junction as the $B$ carrier over the valence-quark picture~\cite{STAR:2024lvy}. This result motivates the inclusion of baryon stopping effects in thermal analyses of isobar systems, which are also incorporated through a factor $\alpha=1/1.84$ in the charge-fraction constraint, \Cref{eq3}, for the theoretical frameworks.  

\begin{figure}[t!]
    \centering
    \includegraphics{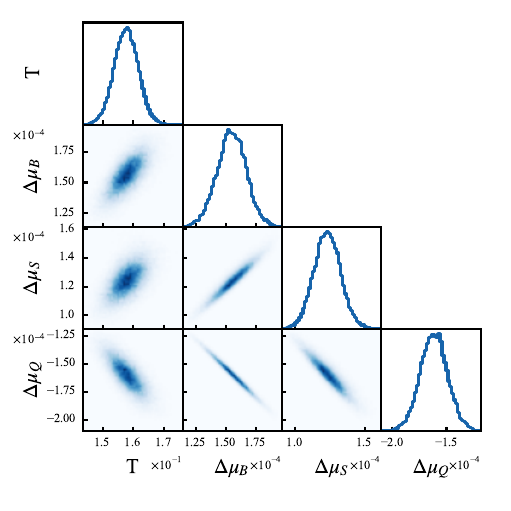}
    \caption{Bayesian analyses posterior showing temperature and chemical potential differences for 0-10\% centrality class. Values in GeV.}
    \label{fig:Baysian}
\end{figure}

In this work, we compare particle yields, yield ratios, and $\Delta Q$ to THERMUS model predictions using a Gaussian-process emulator as a surrogate model within a Bayesian framework. Posterior distributions for the freeze-out parameters are obtained via Markov Chain Monte Carlo (MCMC) sampling, employing uniform priors within physically motivated ranges and a Gaussian likelihood that accounts for both emulator and experimental uncertainties. This procedure yields probability distributions for $T_{\text{Chem}}$, $\mu$'s, and their differences between Ru+Ru and Zr+Zr collisions.

\Cref{fig:Baysian} shows the joint posterior distributions for $T$ and $\Delta \mu_i$, where $\Delta \mu_i=\mu^{Zr}_i-\mu^{Ru}_i$ and $T$ is assumed to be common to both systems. Although uniform priors are employed, their ranges are iteratively refined to exclude regions of negligible posterior weight. Strong correlations are observed among the $\mu$ differences, enabling precise determination of their ratios. Distributions for additional parameters are provided in Supplemental Material, \Cref{B}.
The extracted values of $T$, $\mu_B$, $\mu_S$, and $\mu_Q$ for Ru, together with the corresponding differences and ratios between Ru+Ru and Zr+Zr, are summarized in \Cref{tab:ratios}. Values for Zr+Zr are obtained by adding the extracted differences to the Ru+Ru results. Quantities derived directly from the Bayesian thermal analysis of hadron yields are labeled ``Exp.'' for simplicity. The corresponding predictions from the lattice-based $BQS$ expansion and the mCMF effective model, including their uncertainties, are shown in Supplemental Material, \Cref{C}, where a modest separation between the two isobar systems is observed.

\setlength{\tabcolsep}{.0pt}
\begin{table}[t!]
\centering
\caption{Temperature, chemical potentials, and their differences and ratios from experiment and theory. Values in MeV. $\alpha\neq 1$ includes baryon stopping.}
\label{tab:mus}
\resizebox{\columnwidth}{!}{%
\begin{tabular}{lccccc}
\toprule
 & \multicolumn{5}{c}{Ru+Ru} \\
 \cmidrule(lr){2-6}
 & \multicolumn{2}{c}{$\alpha=1$} & \multicolumn{2}{c}{$\alpha=1/1.84$} & \multicolumn{1}{c}{Exp}  \\
\cmidrule(lr){2-3}\cmidrule(lr){4-5}
 & mCMF & $BQS$ & mCMF & $BQS$ &   \\
\midrule
$T$ & Exp & Exp & Exp & Exp & $158.0_{-3.9}^{+3.9}$ \\
$\mu_{B}$ & Exp & Exp & Exp & Exp & $20.31_{-4.0}^{+3.8}$ \\
$\mu_{S}$ & $3.33_{-0.67}^{+0.67}$ & $4.92_{-1.00}^{+1.00}$ & $3.71_{-0.75}^{+0.75}$ & $5.38_{-1.12}^{+1.13}$ & $3.88_{-3.50}^{+3.30}$ \\
$\mu_{Q}$ & $-0.18_{-0.04}^{+0.04}$ & $-0.23_{-0.09}^{+0.07}$ & $-1.08_{-0.23}^{+0.22}$ & $-1.44_{-0.33}^{+0.31}$ & $0.45_{-3.50}^{+3.60}$ \\
\toprule
 & \multicolumn{5}{c}{Zr+Zr} \\
 \cmidrule(lr){2-6}
 & \multicolumn{2}{c}{$\alpha=1$} & \multicolumn{2}{c}{$\alpha=1/1.84$} & \multicolumn{1}{c}{Exp}  \\
\cmidrule(lr){2-3}\cmidrule(lr){4-5}
 & mCMF & $BQS$ & mCMF & $BQS$ &   \\
\midrule
$T$ &  Exp & Exp & Exp & Exp & $158.0_{-3.9}^{+3.9}$ \\
$\mu_{B}$ & Exp & Exp & Exp & Exp & $20.45_{-4.00}^{+3.80}$ \\
$\mu_{S}$ & $3.43_{-0.67}^{+0.68}$ & $5.05_{-1.00}^{+1.10}$ & $3.78_{-0.75}^{+0.76}$ & $5.47_{-1.12}^{+1.14}$ & $4.00_{-3.50}^{+3.30}$ \\
$\mu_{Q}$ & $-0.37_{-0.08}^{+0.08}$ & $-0.48_{-0.13}^{+0.12}$ & $-1.18_{-0.25}^{+0.24}$ & $-1.58_{-0.36}^{+0.34}$ & $0.29_{-3.50}^{+3.60}$ \\
\toprule
 & \multicolumn{5}{c}{Differentials} \\
 \cmidrule(lr){2-6}
 & \multicolumn{2}{c}{$\alpha=1$} & \multicolumn{2}{c}{$\alpha=1/1.84$} & \multicolumn{1}{c}{Exp}  \\
\cmidrule(lr){2-3}\cmidrule(lr){4-5}
 & mCMF & $BQS$ & mCMF & $BQS$ & \\
\midrule
$\Delta\mu_{B}$  & $0.135_{-5.5}^{+5.5}$ & $0.135_{-5.5}^{+5.5}$ & $0.136_{-5.5}^{+5.5}$ & $0.136_{-5.5}^{+5.5}$ & $0.156_{-0.012}^{+0.011}$ \\
$\Delta\mu_{S}$ & $0.103_{-0.91}^{+0.91}$   & $0.127_{-1.35}^{+1.35}$   & $0.067_{-1.01}^{+1.01}$ & $0.086_{-1.47}^{+1.47}$ & $0.124_{-0.01}^{+0.01}$ \\
\vspace{.1cm}
$\Delta\mu_{Q}$ & $-0.184_{-0.08}^{+0.08}$   & $-0.246_{-0.10}^{+0.10}$  & $-0.103_{-0.31}^{+0.30}$ & $-0.136_{-0.41}^{+0.41}$& $-0.161_{-0.012}^{+0.013}$ \\
\vspace{.1cm}
$\frac{\Delta\mu_{B}}{\Delta\mu_{Q}}$       & $-0.737_{-0.004}^{+0.004}$  & $-0.544_{-0.002}^{+0.002}$  & $-1.301_{-0.004}^{+0.004}$  & $-0.993_{-0.003}^{+0.003}$  & $-0.968_{-0.012}^{+0.011}$ \\
\vspace{.1cm}
$\frac{\Delta\mu_{S}}{\Delta\mu_{Q}}$       & $-0.558_{-0.001}^{+0.001}$  & $-0.512_{-0.0004}^{+0.0004}$   & $-0.641_{-0.001}^{+0.001}$  & $-0.620_{-0.001}^{+0.001}$  & $-0.769_{-0.026}^{+0.021}$ \\
$\frac{\Delta\mu_{B}}{\Delta\mu_{S}}$       & $1.321_{-0.004}^{+0.004}$   & $1.062_{-0.003}^{+0.003}$   & $2.031_{-0.004}^{+0.004}$   & $1.602_{-0.003}^{+0.003}$   & $1.259_{-0.032}^{+0.028}$ \\
\bottomrule
\label{tab:ratios}
\end{tabular}}
\end{table}
To further illustrate the $\mu$ shifts, \Cref{fig:mus_muB_arrows} displays the central values for Ru+Ru and Zr+Zr connected by a red arrow, highlighting both the magnitude and direction of the displacement in the QCD phase diagram. This demonstrates that there is not only a relation among the different $\mu$'s, but also a directionality in the QCD phase diagram showing how $B$ and $S$ vary with $Q$. In \Cref{fig:mus_ratios}, we present the corresponding ratios of $\mu$ differences, $\Delta\mu_i/\Delta\mu_j$, with uncertainties; in some cases, the sign is inverted to allow for a common plotting range.

\subsection{From the theoretical approaches}
\label{Section II_th}

To reproduce the conditions realized in relativistic heavy-ion collisions, theoretical descriptions must satisfy additional constraints. In particular, strangeness neutrality and fixed charge fraction are imposed via
\begin{equation}
    \left<Y_{S}\right>=\left<\frac{n_S}{n_B}\right>=0, \qquad \left<Y_{Q}\right>=\left<\frac{n_Q}{n_B}\right>=\frac{Z}{A}\alpha\,,
    \label{eq3}
\end{equation}
where $n_{B}$, $n_{Q}$, $n_{S}$ are the net baryon, charge, and strangeness densities respectively, and $\alpha=1/1.84$ accounts for baryon stopping effects in 0--10\% most central collisions observed by STAR~\cite{STAR:2024lvy}.  
For both theoretical approaches considered here, the independent input variables are $T$ and $\mu_B$. When combined with the constraints in \Cref{eq3}, these inputs uniquely determine $\mu_S$ and $\mu_Q$ within each theoretical framework, $BQS$ and mCMF. The resulting values for Ru and Zr, with and without additional baryon stopping, are summarized in \Cref{tab:mus} and shown in \Cref{fig:mus_all_models}.

\begin{figure}[t!]
    \centering
    \includegraphics[trim={0 1.46cm 0 0.5cm},clip,width=0.499\textwidth]{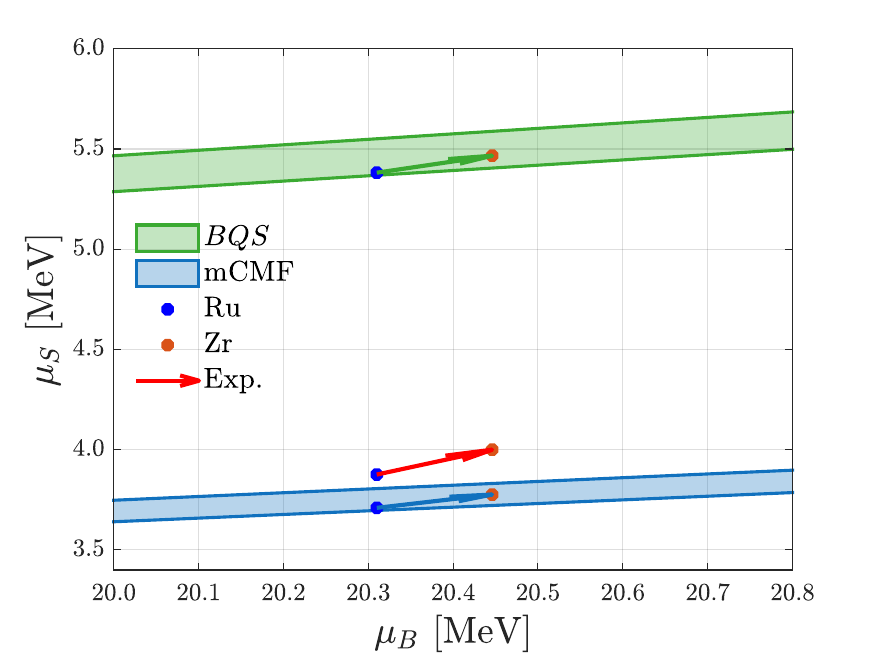}
    \includegraphics[trim={0 0 0 .82cm},clip,width=0.499\textwidth]{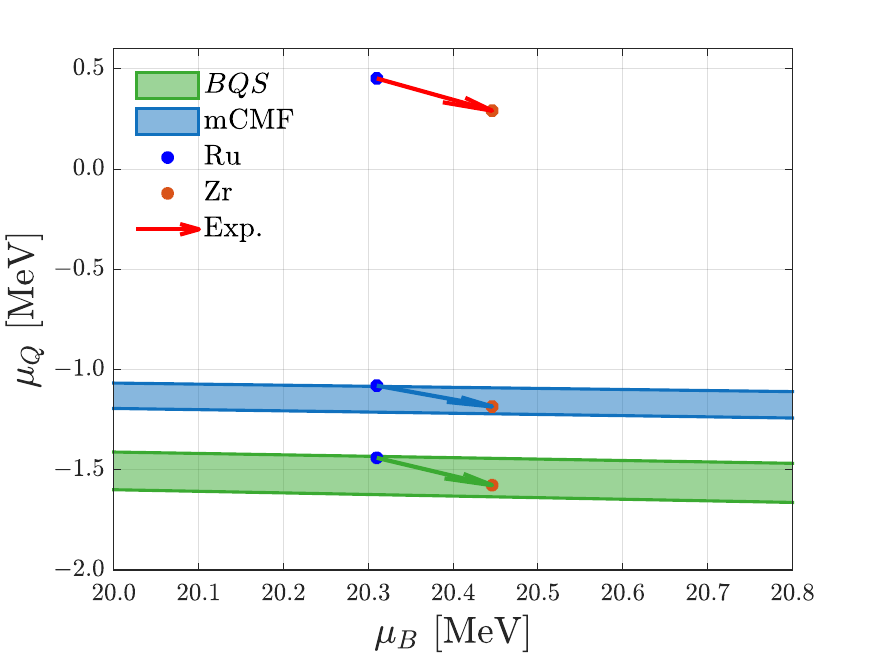} 
    \caption{Strangeness and charge chemical potential change from Ru to Zr from experiment and theory with baryon stopping $\alpha=1/1.84$.}
    \label{fig:mus_muB_arrows}
\end{figure}

Uncertainties in the theoretical $\mu$'s and their differences ($\Delta\mu_B$, $\Delta\mu_S$, and $\Delta\mu_Q$) are then estimated via Monte Carlo propagation. Specifically, the experimental input parameters, $T$ and $\mu_{B}$, are sampled within their quoted uncertainties and propagated through each theoretical framework. The resulting distributions are summarized using the 16th, 50th, and 84th percentiles, and the reported values correspond to the median with asymmetric error bars, written in the form $R_{-a}^{+b}$.
For ratios of $\mu$ differences, namely $\Delta\mu_{B}/\Delta\mu_{Q}$, $\Delta\mu_{S}/\Delta\mu_{Q}$, and $\Delta\mu_{B}/\Delta\mu_{S}$, direct Monte Carlo propagation is not reliable. Because both the numerator and denominator are small and can fluctuate around zero, the resulting ratio distributions may become ill-defined. To obtain statistically meaningful intervals, we instead apply Fieller’s theorem, which provides confidence intervals for ratios of correlated random variables while explicitly accounting for the uncertainty of the denominator. In all cases considered here, the Fieller construction yields finite intervals, allowing the ratios to be reported consistently in the form $R_{-a}^{+b}$.

The differences $\Delta\mu_S$ and $\Delta\mu_Q$ obtained from the theoretical approaches exhibit the same sign as the experimental extractions across all cases shown in \Cref{tab:ratios}. The corresponding ratios likewise agree in sign between theory and experiment. Quantitatively, the $\mu$ differences and ratios from the default mCMF setup (first column) and from $BQS$ with baryon stopping (fourth column) are typically within $\sim30\%$ of the experimental values.

To visualize the $\mu$ shifts, \Cref{fig:mus_muB_arrows} shows the central values of $\mu_B$, $\mu_S$, and $\mu_Q$ for Ru and Zr, connected by arrows for each theoretical approach. The arrows exhibit a consistent direction and similar slopes in both the $\mu_S$–$\mu_B$ (top panel) and $\mu_Q$–$\mu_B$ (bottom panel) planes, reflecting correlated changes among the conserved charges. While the absolute values of $\mu_S$ and $\mu_Q$ from mCMF lie closer to the experimental central values, the slopes of the arrows, corresponding to the $\mu$ shifts, obtained from $BQS$ are generally closer to the experimental trends.

\Cref{fig:mus_muB_arrows} also includes crossover bands obtained using the lattice-QCD pseudocritical line~\cite{Borsanyi:2020fev},
\begin{equation}\label{eq:tran_line}
    \frac{T(\mu_{B})}{T_{c}(\mu_{B}=0)}=1-\kappa_{2}\left(\frac{\mu_{B}}{T(\mu_{B})}\right)^{2}-\kappa_{4}\left(\frac{\mu_{B}}{T(\mu_{B})}\right)^{4}\,,
\end{equation}
where $T_c$ is the pseudocritical temperature at $\mu_B=0$, and $\kappa_2$ and $\kappa_4$ are the Taylor coefficients governing the curvature of the crossover line. This construction is applied consistently within both $BQS$ and mCMF. In both cases, the chemical freeze-out points and arrows lie within the crossover bands, consistent with expectations that chemical freeze-out and chiral crossover occur in close proximity at high collision energies (or low $\mu_B$/baryon density)~\cite{Blaschke:2024jqd,Borsanyi:2025dyp,Koch:2025cog}, here shown as functions of different $\mu$'s (instead of the usual $T$ and $\mu_B$).

\begin{figure}[t!]
    \centering
    \includegraphics[width=0.497\textwidth]{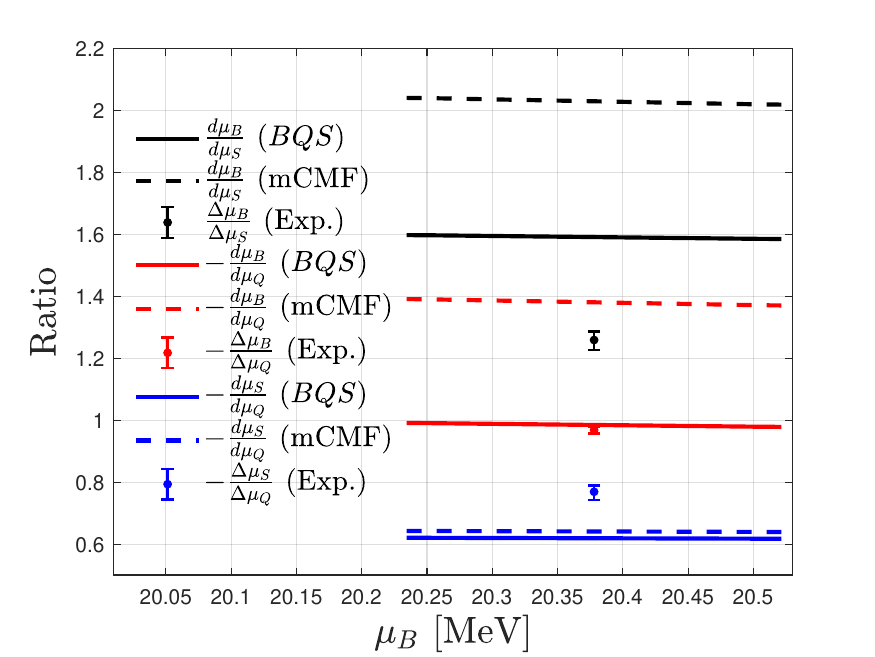}
    \caption{Experimental vs theoretical results for the ratio of chemical potential differences. Results from $BQS$ and mCMF are computed as ($d\mu_{i}/d\mu_{j}$), and experimental results consist of data with error bars ($\Delta\mu_{i}/\Delta\mu_{j}$).}
    \label{fig:mus_ratios}
\end{figure}

Having established that both theoretical approaches reproduce the main experimental trends for isobar collisions, we can further exploit their ability to explore the QCD phase diagram beyond the experimentally accessible region. In particular, the theoretical frameworks allow the computation of continuous derivatives $d\mu_i/d\mu_j$ as functions of $\mu_B$. \Cref{fig:mus_ratios} compares these derivatives with the experimentally extracted  ratios. Notably, the $B$ to $Q$ ratio shows the closest agreement between experiment and the $BQS$ expansion, whereas the $S$ to $Q$ ratio is better reproduced by mCMF. The largest deviations occur in the $B$ to $S$ channel, where $BQS$ nevertheless remains closer to the experimental values than mCMF.

\section{Discussion and conclusions}
\label{Section III}

STAR measurements of isobar collisions at $\sqrt{s_{NN}}=200$~GeV populate slightly different freeze-out points for Ru and Zr in the multidimensional QCD phase diagram. 
We perform a Bayesian thermal analysis that fits the two systems simultaneously, using identified-hadron yields and the net-charge yield differences between isobars, rather than fitting Ru and Zr independently. This strategy substantially reduces systematic uncertainties and enables a precise extraction of the chemical-potential differences ($\Delta\mu_i\equiv \mu_i^{\rm Zr}-\mu_i^{\rm Ru}$), even when the individual chemical potentials, particularly ($\mu_Q$) and ($\mu_S$), carry comparatively larger uncertainties.

The resulting values of $T_{\rm Chem}$, $\mu_i$, and $\Delta\mu_i$ are summarized in \Cref{tab:mus}. These results demonstrate that small changes in $Y_Q$ generate measurable shifts in freeze-out conditions and provide a precision probe of correlations among conserved charges.

To confront these experimental constraints with theoretical frameworks, we impose strangeness neutrality ($Y_S=0$) and fix $Y_Q$ to the experimentally determined isobar values. Using as inputs the Bayesian $T_{\text{Chem}}$ and $\mu_B$ (and including baryon stopping when $\alpha\neq1$), we compute $\mu_S$ and $\mu_Q$ within the Taylor-expanded lattice-QCD equation of state ($BQS$)~\cite{Noronha-Hostler:2019ayj} and within the effective mCMF model~\cite{Kumar:2025rxj}. As shown in \Cref{tab:mus}, both approaches reproduce the sign and overall scale of the extracted $\mu_S$ and $\mu_Q$ for Ru and Zr, and yield $\Delta\mu_S$ and $\Delta\mu_Q$ that agree with experiment in sign and approximate magnitude. These patterns are driven primarily by the neutron excess of the colliding nuclei, which sets the observed charge and strangeness response.

The experimentally extracted differences can be cast into ratios $\Delta\mu_i/\Delta\mu_j$, which quantify the correlated response of the chemical potentials. 
The observed negative signs of $\Delta \mu_{B}/\Delta \mu_{Q}$ and $\Delta \mu_{S}/\Delta \mu_{Q}$ follow from isospin imbalance in neutron rich matter and mirror the sign of the corresponding derivatives $d\mu_B/d\mu_Q$ and $d\mu_S/d\mu_Q$ in the theoretical frameworks. Experiment and theory agree on the sign of all ratios, with the closest quantitative agreement occurring for $\Delta\mu_B/\Delta\mu_Q$ in $BQS$ and for $\Delta\mu_S/\Delta\mu_Q$ in mCMF, while $\Delta\mu_B/\Delta\mu_S$ shows the largest deviations. Beyond these discrete experimental ratios, the theoretical frameworks provide continuous derivatives $d\mu_i/d\mu_j$ and trajectories throughout the phase diagram.

This advantage becomes particularly clear when comparing the isobar freeze-out region ($\mu_B\simeq20$~MeV) to the $\mu_B\to0$ limit. At freeze-out, the ratios are modest, $\Delta\mu_B/\Delta\mu_Q\approx-1$, $\Delta\mu_S/\Delta\mu_Q\approx-0.77$, and $\Delta\mu_B/\Delta\mu_S\approx1.26$, and both $BQS$ and mCMF yield similar values. In contrast, at $\mu_B=0$ the theoretical derivatives are larger in magnitude. For $BQS$ we find $d\mu_B/d\mu_Q\approx-12.25$, $d\mu_S/d\mu_Q\approx-3.33$, and $d\mu_B/d\mu_S\approx3.67$, while mCMF gives $d\mu_B/d\mu_Q\approx-16$, $d\mu_S/d\mu_Q\approx-3$, and $d\mu_B/d\mu_S\approx5.35$. This strong variation over a change of only $\sim20$~MeV in $\mu_B$ highlights the enhanced sensitivity of the response near vanishing net baryon density.

At $\mu_B=0$ and $Y_S=0$, the derivative $d\mu_B/d\mu_S$ can be related to the baryon-strangeness correlation,
\begin{equation}
C_{BS}(T,\mu_B,\mu_S)=-3\,\frac{\chi_{11}^{BS}(T,\mu_B,\mu_S)}{\chi_{2}^{S}(T,\mu_B,\mu_S)}\,,
\end{equation}
with generalized susceptibilities defined by
\begin{equation}
\chi_{ij}^{BS}(T,\mu_B,\mu_S)=\frac{\partial^{i+j}P(T,\mu_B,\mu_S)/T^4}{\partial(\mu_B/T)^i\,\partial(\mu_S/T)^j}\,,
\label{Eq6}
\end{equation}
 as chemical-potential derivatives of the pressure $P$. This implies $d\mu_B/d\mu_S=3/C_{BS}$ under $Y_S=0$ at $\mu_B=0$, consistent with previous lattice studies of strangeness-neutral QCD thermodynamics~\cite{Koch:2005vg,Fu:2018swz}. In the confined phase, $\chi_{11}^{BS}$ receives contributions from strange (anti)baryons, while $\chi_2^S$ also includes strange mesons; as $\mu_B$ increases (and $T_{\rm Chem}$ decreases), baryonic contributions become relatively more important, increasing $C_{BS}$ and thus reducing $d\mu_B/d\mu_S$.

Neither theoretical approach uniformly outperforms the other. mCMF is calibrated to reproduce lattice thermodynamics at high $T$ and low $\mu_B$ for $\mu_Q=\mu_S=0$~\cite{Kumar:2025rxj}, while its isospin dependence is constrained primarily at $T=0$ by nuclear and astrophysical observables~\cite{Dexheimer:2008ax,Kumar:2024owe}. This dual calibration illustrates the connection between heavy-ion collisions and neutron-star matter across widely separated regions of the phase diagram. Residual discrepancies, most prominently in $\Delta\mu_B/\Delta\mu_S$, may indicate missing strange and non-strange resonances in the effective description; extensions along these lines are underway.

In summary, we have presented the first extraction of the $\mu$'s and $T_{\text{Chem}}$ in relativistic heavy-ion collisions from a full Bayesian analysis of a thermal fit to hadron yields and ratios, enabled by a novel method that isolates the net $Q$ of the system through yield differences between isobar species. This approach substantially reduces systematic uncertainties and allows, for the first time, a precise determination of $T_{\text{Chem}}$, the individual $\mu$'s, and their differences between Ru+Ru and Zr+Zr collisions. Since $\sqrt{s_{NN}}=200$ GeV is the only energy at which isobar collisions have been performed, these results uniquely demonstrate the power of controlled isospin variations to probe the multidimensional QCD phase diagram with high precision. By combining experimental measurements with a lattice-QCD-based expansion and an effective model, our work establishes a framework for connecting heavy-ion phenomenology with first-principles QCD and astrophysical constraints, paving the way toward unified insights across the QCD phase diagram.


\begin{acknowledgments}
This work is partially supported by the NP3M Focused Research Hub supported by the National Science Foundation (NSF) under grant No. PHY-2116686. We also acknowledge support by the NSF under grants No. PHY-2208724, and PHY-2514763, and within the framework of the MUSES collaboration, under grant number No. OAC-2103680. This material is also based upon work supported by the U.S. Department of Energy, Office of Science, Office of Nuclear Physics, under Award Number DE-SC0022023 and DE-FG02-89ER40531, and by the National Aeronautics and Space Agency (NASA)  under Award Number 80NSSC24K0767.

\end{acknowledgments}
\bibliographystyle{apsrev4-2}
\bibliography{references}

\pagebreak
\widetext
\begin{center}
\textbf{\large Supplemental material}
\end{center}
\setcounter{equation}{0}
\setcounter{figure}{0}
\setcounter{table}{0}
\setcounter{section}{0}
\setcounter{page}{1}
\makeatletter
\renewcommand{\theequation}{S\arabic{equation}}
\renewcommand{\thefigure}{S\arabic{figure}}
\renewcommand{\thetable}{S\arabic{table}}
\renewcommand{\thesection}{S\arabic{section}}

\section{Estimation of particle yields }
\label{A}

The $p_{T}$ spectra of pions, kaons and protons for the isobar systems are available in Ref~\cite{STAR:2024lvy}. The integrated yields used in this work are obtained by fitting the published spectra with a blast‑wave (BW) function~\cite{Schnedermann:1993ws} separately for each particle species. The measured data points in the $p_{T}$ region where data exist are summed directly, while the contribution from the unmeasured $p_{T}$ region is estimated by integrating the extrapolated BW fit over that range. Central values for ratios are calculated from yields directly. 

Systematic uncertainties arising from the extrapolation are taken into account. Each spectrum is fitted using three different functional forms:  
\begin{enumerate}
\item Bose–Einstein distribution,  
\item Lévy function, and  
\item BW with a reduced upper $p_{T}$ limit.  
\end{enumerate}
The Barlow method~\cite{Barlow:2002yb} is employed to estimate the uncertainty associated with (i) the choice of fit function (cases 1 and 2) and (ii) the restricted fit range (case 3). The uncertainties from these two sources are combined in quadrature. 
Besides, additional global systematic uncertainties of 2\%, 3\% and 5\% are assigned to the $\pi^{-}/\pi^{+}$ ratio, the $K^{-}/K^{+}$ ratio, and all other yields and ratios, respectively. These global uncertainties represent uncertainties in tracking efficiencies and are consistent with previous STAR analyses~\cite{STAR:2008med}.

\section{Prior and full posterior distributions \label{B}}

The priors for the Bayesian analysis parameters used in this work are uniform distributions over the ranges listed in~\Cref{tab:paraRange}. These ranges are selected iteratively, beginning with wide bounds for all parameters and progressively narrowing them by excluding regions of parameter space that yield essentially zero posterior probability. After each refinement, the emulator is retrained using only training data within the updated priors. Training on a reduced parameter space lowers interpolation uncertainty because training points are closer to each other. This process is repeated until the interpolation uncertainty falls below the level of the experimental data uncertainty.

\begin{table}[h]
    \caption{List of free parameters for THERMUS and ranges used in the Bayesian analysis.}
    \label{tab:paraRange}
    \centering
    \begin{tabular}{cc}
        \hline
        Parameter & Range \\
        \hline
        $T_{\text{Chem}}$* & 0.13 to 0.19 GeV \\
        $\mu_B(\mathrm{Ru})$ & $-0.05$ to $0.05$ GeV \\
        $\mu_S(\mathrm{Ru})$ & $-0.01$ to $0.01$ GeV \\
        $\mu_Q(\mathrm{Ru})$ & $-0.02$ to $0.02$ GeV \\
        $\gamma_S(\mathrm{Ru})$ & 0.3 to 1.0 \\
        $\Delta\mu_B$ & $-0.03$ to $0.03$ GeV \\
        $\Delta\mu_S$ & $-0.003$ to $0.003$ GeV \\
        $\Delta\mu_Q$ & $-0.002$ to $0.002$ GeV \\
        $R(\mathrm{Ru})$ & 0 to 7.5 fm \\
        $\Delta R$ & $-1$ to $1$ fm \\
        \hline
        \multicolumn{2}{l}{\footnotesize *Temperatures for Ru and Zr are assumed identical at chemical freeze-out. $\Delta\mu_{i}=\mu_{i}^{Zr}-\mu_{i}^{Ru}$}
    \end{tabular}
\end{table}

\begin{figure*}
    \centering
    \includegraphics[width=0.8\textwidth]{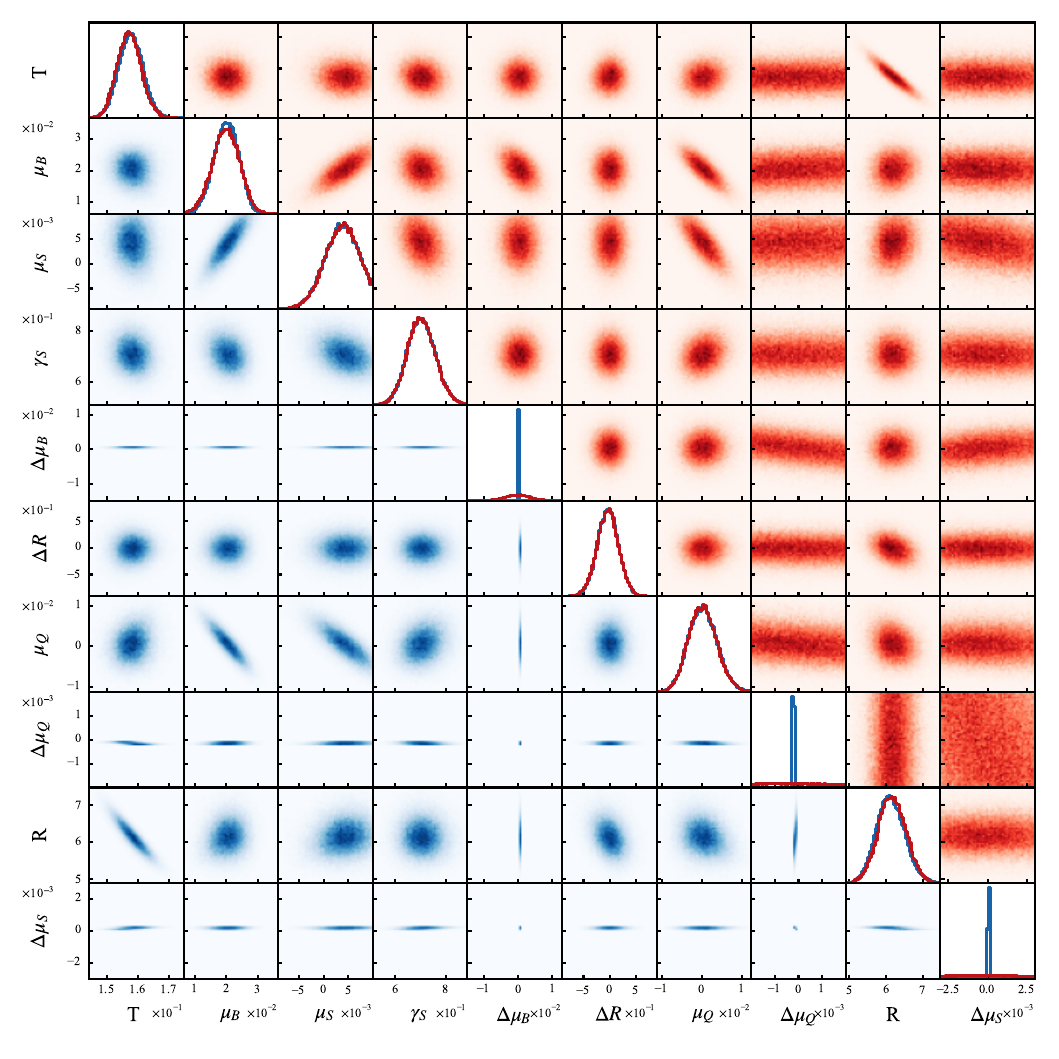}
    \caption{Bayesian analysis posterior distributions for all pairs of parameters. Energies are expressed in GeV for $T$, $\mu_{B}$, $\mu_{S}$, $\Delta\mu_{S}$ and $\Delta\mu_{Q}$ ($\gamma_{S}$ is omitted as being dimension-less), and distances are in fm for $R$ and $\Delta R$. The lower‑triangular half of the matrix shows the posterior when $\Delta Q$ is included in the THERMUS fit, while the upper‑triangular half displays the posterior obtained without $\Delta Q\,. $ The diagonal histograms show the marginalized probability distributions both with $\Delta Q$ (light blue) and without $\Delta Q$ (dark red).}
    \label{fig:fullBayesian}
\end{figure*}

The full joint posterior distributions involving all THERMUS parameters, i.e. $T$, $\mu_{B}$, $\mu_{S}$, $\gamma_{S}$, $\Delta\mu_{S}$, $\Delta\mu_{Q}$, $R$, and $\Delta R$, are shown in \Cref{fig:fullBayesian}. The lower‑left rectangle shows the posterior when THERMUS is fitted to particle yields, yield ratios, and $\Delta Q_i$ for $i\in\{\pi,K,p\}$. The upper‑right rectangle shows the posterior when $\Delta Q_i$ is omitted from the fit. This demonstrates that no constraints on any of the $\Delta\mu$ parameters are possible without fitting $\Delta Q_i$, a measurement that can only be performed with the isobar data set. The diagonal histograms display the marginalized probability distributions both with $\Delta Q_i$ (light blue) and without $\Delta Q_i$ (dark red).

\begin{figure*}
    \centering
        \includegraphics[trim={0 .75cm 0 0}, clip,width=.33\textwidth]{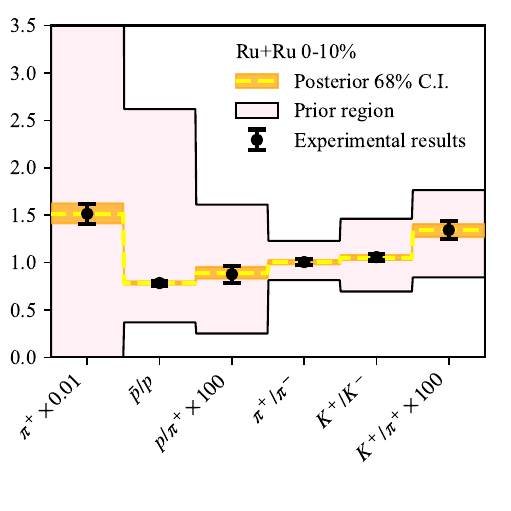}
        \includegraphics[trim={0 .75cm 0 0}, clip,width=.33\textwidth]{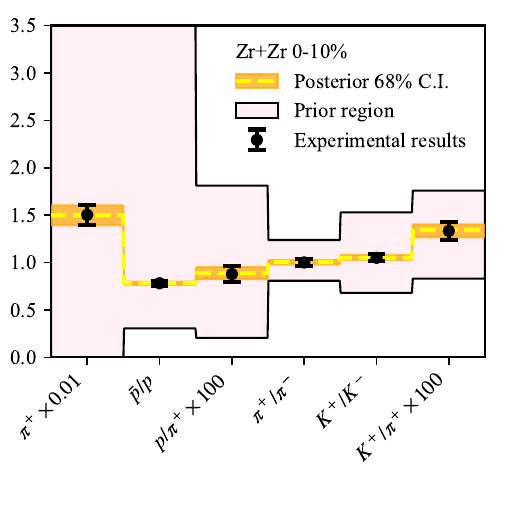}
        \includegraphics[trim={0 0 0 0}, clip,width=.32\textwidth]{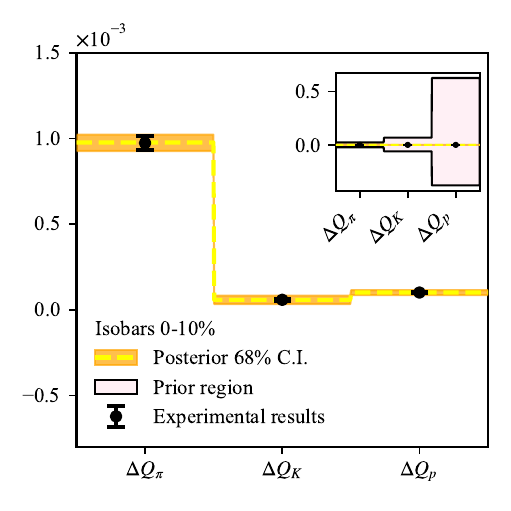}
    \caption{Prior ranges and posterior predictions for the observable quantities, with experimental data plotted as black markers. The left and middle panels show particle yields and ratios for Ru+Ru and Zr+Zr, while the right panel shows net-charge yield differentials $\Delta Q$s. The pink region indicates the span of observables allowed by the priors, while the yellow dashed line represents the median values from the posterior distribution. The $\Delta Q$ plot inset improves legibility because the prior range for $\Delta Q_{p}$ is far off‑scale. Pion yields and $\Delta Q$s given in GeV$^{-2}$c$^{2}$.}
    \label{fig:THERMUSVsData}
\end{figure*}

The agreement between data and the fit is shown in  \Cref{fig:THERMUSVsData}. The pink patch in the background shows the range of all possible parameter values given by the range of priors. The 68\% posterior confidence intervals (C.I.) agree well with experimental results. Our choice of particle yields and ratios is labeled on the $y$-axis. Particle ratios are used to partially cancel out the systematic uncertainties in the estimation of particle yields. 

\section{Additional plots for chemical potentials and theory comparison}
\label{C}

\begin{figure*}[t!]
    \centering
    \includegraphics[width=0.497\textwidth]{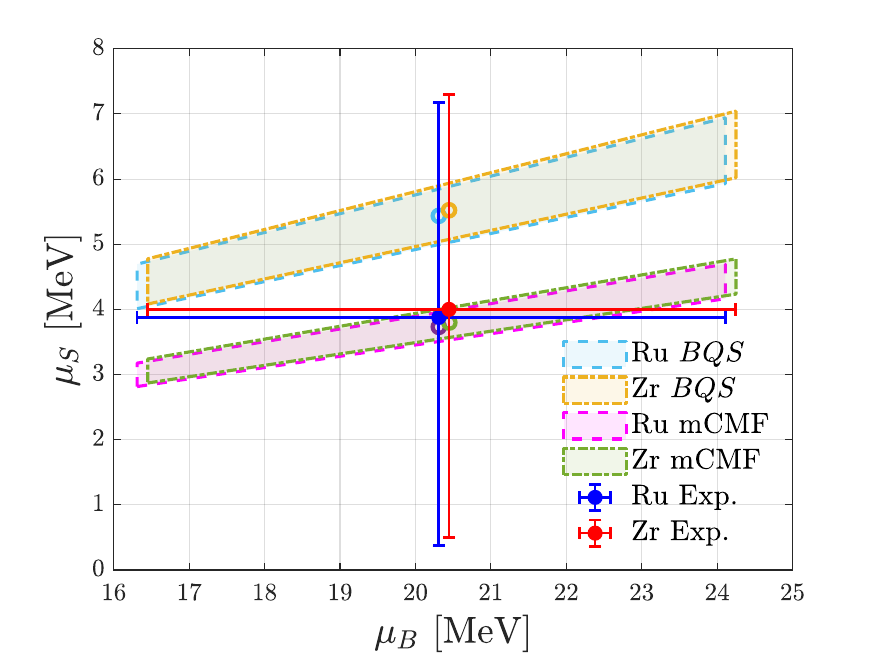}
    \includegraphics[width=0.497\textwidth]{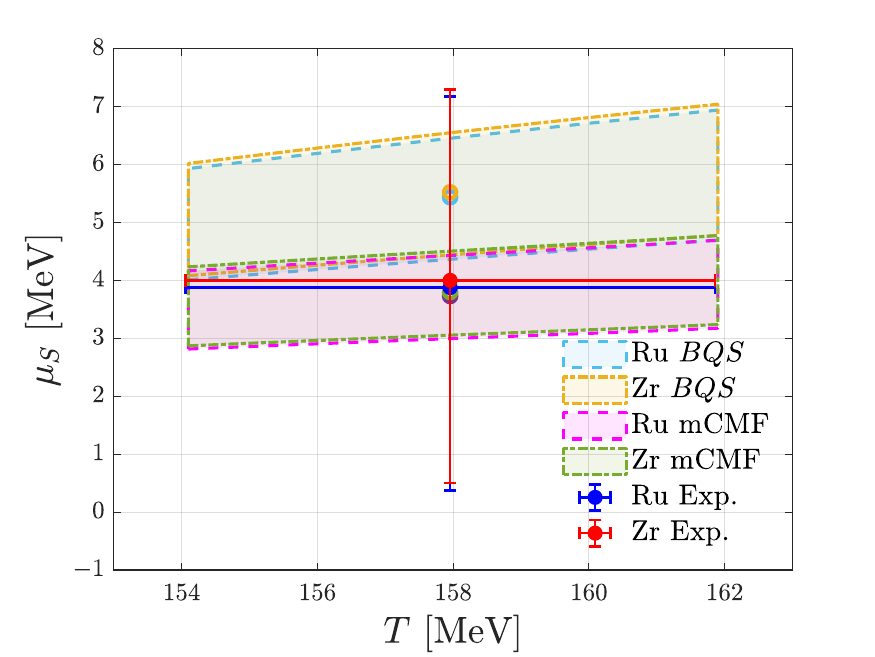}
    \includegraphics[width=0.497\textwidth]{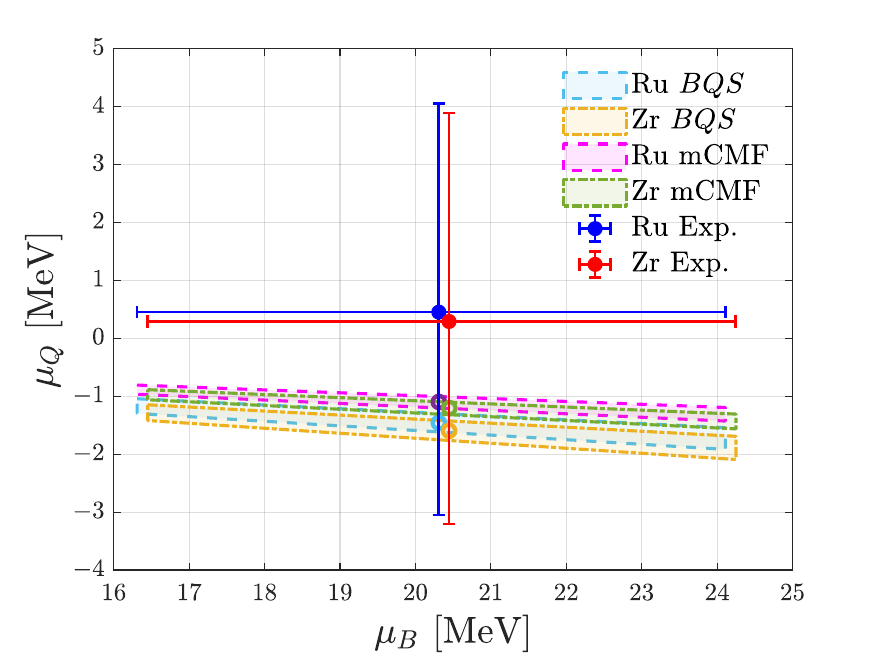}
    \includegraphics[width=0.497\textwidth]{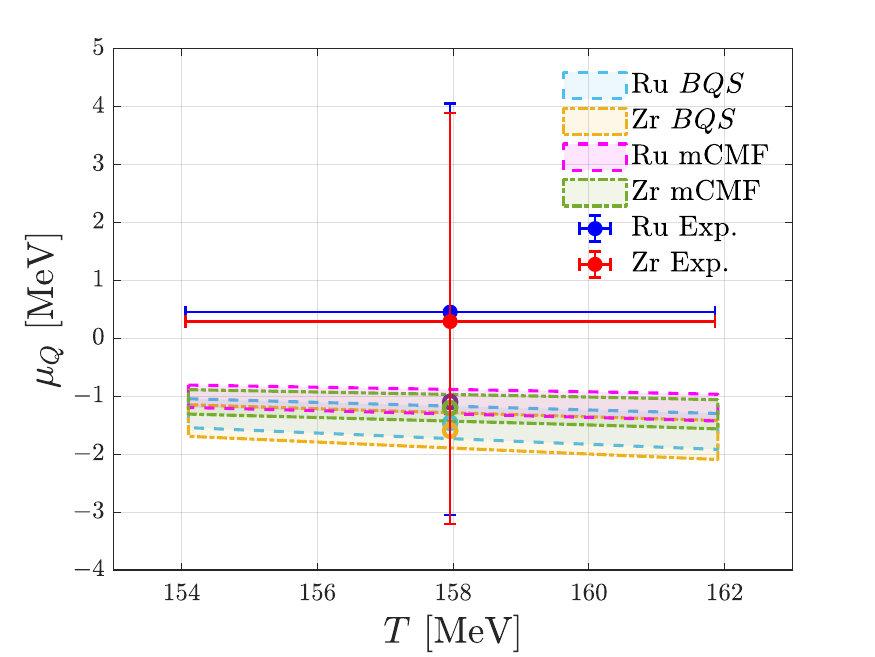}
   \caption{ The strangeness, baryon, and charged chemical potentials and temperature from the theoretical frameworks (including baryon stopping $\alpha=1/1.84$) and experiment for Ru and Zr from the hadronic yields THERMUS fitting within the Bayesian framework. Values are provided in \Cref{tab:ratios}.}
    \label{fig:mus_all_models}
\end{figure*}

\Cref{fig:mus_all_models} illustrates the values shown in \Cref{tab:mus} for Ru and Zr with baryon stopping $\alpha\neq 1$. The filled circles and lines (with error) show the experimental values, while the empty circles and regions (with propagated error up to 1$\sigma$) show results for $BQS$ and mCMF. Across the different panels, with different $T$ and $\mu$ relations, mCMF results are consistently slightly closer to experiment than $BQS$, with better agreement for the upper panels (that involve $\mu_S$) than the lower panels (that involve $\mu_Q$).

\end{document}